\documentclass[sigconf,screen]{acmart}
\AtBeginDocument{%
  \providecommand\BibTeX{{%
    \normalfont B\kern-0.5em{\scshape i\kern-0.25em b}\kern-0.8em\TeX}}}

\setcopyright{acmcopyright}
\copyrightyear{2018}
\acmYear{2018}
\acmDOI{XXXXXXX.XXXXXXX}

\usepackage{wrapfig}
\usepackage{multirow}
\usepackage[shortlabels]{enumitem}
\setlist{topsep=2pt, itemsep=2pt}
\usepackage[compact]{titlesec}
\usepackage[bottom]{footmisc}
\usepackage{float}
\usepackage{subcaption}
\usepackage{graphicx}
\usepackage{lipsum}
\usepackage{fdsymbol}
\usepackage{amsmath}
\usepackage[belowskip=-12pt,aboveskip=0pt]{caption}
\usepackage{caption}

\begin{document}

\title{Fairness Through Domain Awareness: Mitigating Popularity Bias For Music Discovery}


\author{Rebecca Salganik}
\affiliation{%
  \institution{Universite de Montreal/MILA}
  \city{Montreal}
  \state{Quebec}
  \country{Canada}
}
\email{rebecca.salganik@umontreal.edu}

\author{Fernando Diaz}
\affiliation{%
  \institution{Google/MILA}
  \city{Montreal}
  \state{Quebec}
  \country{Canada}
}
\email{diazf@acm.org}

\authornotemark[1]
\author{Golnoosh Farnadi}
\authornotemark[1]
\affiliation{%
  \institution{HEC/Universite de Montreal/MILA}
  \city{Montreal}
  \state{Quebec}
  \country{Canada}
}
\email{farnadiq@mila.quebec}

\begin{abstract}

As online music platforms grow, music recommender systems play a vital role in helping users navigate and discover content within their vast musical databases. At odds with this larger goal, is the presence of popularity bias, which causes algorithmic systems to favor mainstream content over, potentially more relevant, but niche items. In this work we explore the intrinsic relationship between music discovery and popularity bias. To mitigate this issue we propose a domain-aware, individual fairness-based approach which addresses popularity bias in graph neural network (GNNs) based recommender systems. Our approach uses individual fairness to reflect a ground truth listening experience, i.e., if two songs sound similar, this similarity should be reflected in their representations. In doing so, we facilitate meaningful music discovery that is robust to popularity bias and grounded in the music domain. We apply our BOOST methodology to two discovery based tasks, performing recommendations at both the playlist level and user level. Then, we ground our evaluation in the cold start setting, showing that our approach outperforms existing fairness benchmarks in both performance and recommendation of lesser-known content. Finally, our analysis explains why our proposed methodology is a novel and promising approach to mitigating popularity bias and improving the discovery of new and niche content in music recommender systems.


\end{abstract}





\maketitle

\section{Introduction}

The proliferation of market activity on digital platforms has acted as a catalyst for research in recommendation, search, and information retrieval \cite{hossain_survey_2023}. At its core, the goal of this research is to design systems which can facilitate users' exploration of an extensive item catalogue: be it in the domain of journalism \cite{wu_personalized_2022}, films \cite{harper_movielens_2015}, fashion \cite{ding_personalized_2023}, music \cite{salha-galvan_GAE_2021, korzeniowski_artist_2021, saravanou_multi-task_2021}, or otherwise. Within this larger goal of recommendation, each domain comes with its own specifics that differentiate it from other settings \cite{noh_study_2023, ekstrand_user_2014, burke_matching_2011}. Particular to the music streaming domain, an extensive body of work has explored the importance of discovery, exploration, and novelty in the larger goal of performing music recommendation \cite{drott_why_2018, lavranos_theoretical_2016, cunningham_finding_2007, raff_music_2021, mantymaki_gratifications_2015, gathright_understanding_2018}. Broadly, discovery can be considered the ability of a curatorial system to expose users to relevant content that they would not have manually discovered themselves  \cite{raff_music_2021, herlocker_evaluating_2004, gathright_understanding_2018}. And, most significantly, a collection of works have shown that music discovery to be the second most important factor for customer loyalty respective to a particular streaming platform due to the gratifying nature of constructing playlists and interacting with an algorithmic curatorial system  \cite{mantymaki_gratifications_2015, lavranos_theoretical_2016, raff_music_2021}. 

Crucially, recent work in this domain has begun to uncover an inverse relationship between novelty, one of the keys to discovery, and the notion of popularity bias \cite{wu_personalized_2022, kamehkosh_user_2017}. Within the broader recommendation community, popularity bias has long been an important topic of research. This phenomenon manifest itself when algorithmic reliance on pre-existing data causes new, or less well known items, to be disregarded in favor of previously popular items \cite{jannach_what_2015, park_long_2008, steck_item_2011, chen_bias_2020, abdollahpouri_connection_2020, celma_from_2008}. And, particularly in the context of discovery, where purpose of a user's engagement with algorithmic curation hinges on exposure to musical items which they would not have already been familiar with, the presence of popularity bias can clearly hinder a system's ability to serve this need. In this work, we explore the \textbf{intrinsic interplay between discovery and popularity bias} through the lens of graph neural network (GNN) based recommender systems \cite{wu_graph_2022, gao_survey_2023}. In the graph space, popularity of individual items is deeply interlaced with the degree centrality of a node, or the number of outgoing edges that leave this node and connect it to others in the graph. This is because the innate process of representation learning is affected by the number of neighbors a node has \cite{kang_rawls_2022}. And, thus, a node's centrality can dictate the quality of its learned representation. This suggests that duplicating the feature information of an extremely popular song, creating a new song using these duplicate features, and randomly placing it once at the edge of a graph, will significantly impact its learned representation, even if everything about the song remains \textit{exactly the same}. As we show in our experimentation, one solution to this disparity lies in a debiasing method that is aware of similarities between musical items and is, thus, grounded in the musical domain. 

However, current approaches for mitigating popularity bias in recommender systems approach this task in a domain agnostic approach \cite{rhee_countering_2022, abdollahpouri_managing_2019, mansoury_graph_2021, zhang_incorporating_2023, wei_model_2021}. Such abstraction can be extremely relevant to domains in which item and user level features are scarce, sparse, or non existent. However, in an environment like music streaming where there is a plethora of valuable feature information, we believe that grounding fairness notions in domain specific attributes can prove incredibly valuable. In addition, a majority of these methods focus on using either group \cite{rhee_countering_2022, zhang_incorporating_2023} or counterfactual fairness \cite{zhao_investigating_2022, zhu_popularity_2021}, often relying on a binary sensitive attribute to encode popularity. This can cause intrinsic limitations because popularity between items is not necessarily a binary state and such attributes may not be readily available.

In this work, we propose a domain aware, individual fairness based approach for facilitating engaging music discovery. In order to facilitate the domain awareness of our approach we generate nuanced multi-modal track features, extensively augmenting two publicly available datasets. Using these novel feature sets, we show the importance of integrating musical similarity into a debiasing technique and show the effects of our method at learning expressive representations of items that are robust to the effects of popularity bias in the graph setting. Grounding our approach in the musical domain empowers us to leverage a ranking-based individual fairness definition and extend it to the bipartite graph setting. In doing so, we design a method that uses music features to fine-tune item representations such that they are reflective of information that is, in essence, a ground truth to the listening experience: two songs that sound similar should, at least somewhat, reflect this similarity in their learned representations. Finally, we compare our individual fairness-based method with three other methods which are grounded in other canonical fairness notions and are not domain-aware. Through a series of empirical results, we show that our fairness framework enables us to outperform a series of accepted fairness benchmarks in both performance and recommendation of lesser known content on two important music recommendation tasks. In summary, the contributions of this paper are the following: 
\begin{enumerate}
    \item \textbf{Problem Setting}: we define the task of music discovery through the lens of domain-aware individual fairness, showing the intrinsic connections between individual fairness, musical similarity, popularity bias, and music discovery. 
    \item \textbf{Dataset Design}: we extensively augment two classic music recommendation datasets to generate a set of nuanced multi-modal track features, laying the foundation for future domain-aware mitigation techniques. 
    \item \textbf{Method}: (1) we provide a novel technical formulation of popularity bias (2) design a domain-aware ranking based individual fairness approach to mitigating popularity bias in graph-based recommendation. 
    \item \textbf{Experiments}: we show that our method outperforms three state of the art fairness benchmarks in the cold start setting.
\end{enumerate}
\raggedbottom

\section{Related Work}
In this section we contextualize our work by presenting relevant literature on the subjects of (1) popularity bias and (2) graph neural network (GNN) based recommender systems. 
\raggedbottom
\subsection{Popularity Bias in Recommendation} \label{sec:pop_def_rw}
\begin{figure}
  \begin{center}
    \includegraphics[scale=0.15]{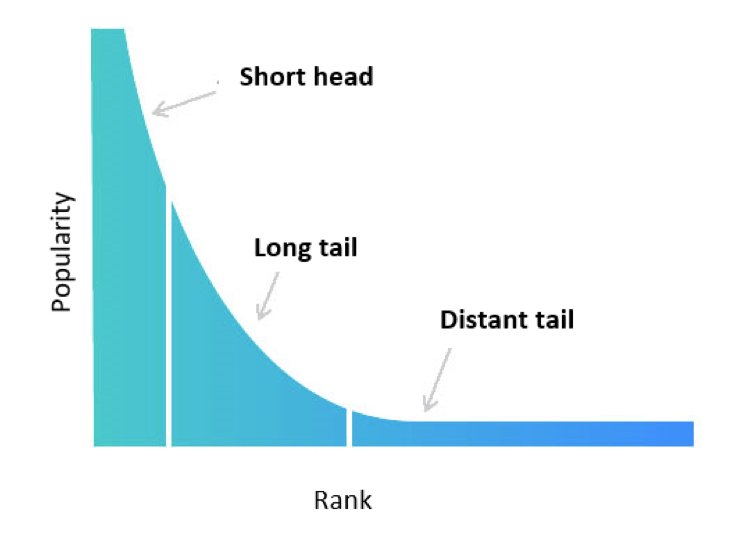}
  \end{center}
  \caption{Classic definition of Long Tail popularity settings\cite{abdollahpouri_controlling_2017}}
  \label{classic_LT}
\end{figure}

Most broadly, popularity bias refers to a disparity between the treatment of popular and unpopular items at the hands of a recommender system. As such, this term is loosely tied to a collection of complementary terms including exposure bias \cite{Diaz__evaluating_2020}, superstar economics \cite{bauer_music_2017}, long tail recommendation \cite{mansoury_graph_2021}, the Matthew effect \cite{moller_not_2018}, and aggregate diversity \cite{adomavicius_improving_2012, celma_novel_2008}. 
There have been several different approaches to formulating popularity through some quantitative definition. One body of work defines popularity with respect to individual items' visibility \cite{Zhang_causal_2021, Diaz__evaluating_2020, mansoury_graph_2021}. Another group of approaches attempts to simplify this process by applying some form of binning to the raw appearance values. Most notably, the long tail model \cite{celma_from_2008, goel_anatomy_2010, downey_evidence_2001, yang_visualization_2003, park_long_2008} has risen to prominence as a popularity definition. As shown in Figure \ref{classic_LT}, we can see that the first 20\% of items, called \textit{short head}, take up a vast majority of the user interactions and the remaining 80\%, or \textit{long tail} and \textit{distant tail}, have, even in aggregate, significantly fewer interactions. Often, splitting items into the \textit{short head} and \textit{long tail} (either including or removing \textit{distant tail}) to define disparity in popularity can overcome the issues of range while still representing concrete disparities in item level visibility.

There has been a lot of work done analyzing and codifying the nature of popularity bias in recommender systems \cite{canmares_should_2018, Jadidinejad2019HowSI, jannach_what_2015, chen_bias_2020}. Approaches to mitigating popularity bias are often grounded in one of three methods: \textbf{pre-processing} \cite{boratto_interplay_2021}, \textbf{in-processing} \cite{rhee_countering_2022, wei_model_2021, zheng_2021_disentangling}, or \textbf{post-processing}, \cite{abdollahpouri_managing_2019, mansoury_graph_2021}. These mitigation strategies are often based on the instrumentation of various canonical fairness notions such as \textbf{group fairness} \cite{boratto_interplay_2021, schnabel2016recommendations, rhee_countering_2022, zhang_incorporating_2023, abdollahpouri_managing_2019}, \textbf{counterfactual fairness} \cite{wei_model_2021, zheng_2021_disentangling, Zhang_causal_2021}, or \textbf{individual fairness} \cite{Chakraborty2017FairSF, wang_providing_2022}. 

We contrast our work with previous individual fairness approaches in our use of the music feature space as a form of domain expertise in definition of item-item similarity. We argue that without this ``anchoring'' an individual fairness method that uses the output of a recommender model, whether it be in learned representation \cite{wang_providing_2022} or the relevance score \cite{Chakraborty2017FairSF}, is already influenced by an item's popularity. Finally, in addition to the classical formulation of popularity bias, a group of works have explored the connection between popularity bias and novelty \cite{lo_matching_2019, zhao_investigating_2022, zhu_popularity_2021} where various metrics are designed to evaluate the novelty of a recommended list. We see our work as complimentary to the exploration in this area however, we differentiate our problem formulation because while novelty is an important aspect of discovery, without domain awareness novelty alone does not account for musical similarity - a critical aspect of the discovery setting.

\subsection{GNNs in Recommendation} \label{sec:GNN_in_rec}
In recent years various graph neural network (GNN) architectures have been proposed for the recommendation domain \cite{wu_graph_2020}. For brevity, we will focus only on the two methods that are used as the backbone recommenders to the fairness mitigation techniques discussed later in this paper, however we refer to the following surveys \cite{wu_graph_2022, gao_survey_2023} for recent innovations in this domain. 

In particular, \emph{PinSage} \cite{ying_graph_2018} is an industry solution to graph-based recommendation. Unlike many competing methods, which train on the entire neighborhood set of a node, PinSage trains on a randomly sampled subset of the graph. In order to construct neighborhoods, PinSage uses $k$ random walks to select the top $m$ most relevant neighbors to use as the neighbor set. However, it is important to note that PinSage learns representations of items but not users. Meanwhile, \emph{LightGCN} \cite{he_lightgcn_2020}, is a method that learns both user and item embeddings simultaneously. Since its proposal in 2020, it is still considered state of the art. 



\section{Methodology}
In this section we detail the dataset augmentation procedure and architecture of our domain-aware, individually fair music recommendation system. First, we introduce our datasets in Section~\ref{sec:data_aug}. Then, following the problem setting in Section~\ref{sec:prob_settings}, we reformulate popularity bias in Section~\ref{sec:pop_def} and  introduce our domain-aware, individually fair music recommender system in Section~\ref{sec:r_arch}.

\subsection{Dataset Augmentation Procedure} \label{sec:data_aug}

One of the limitations with working on music recommendation is the scarcity of up-to-date, publicly available feature-based datasets. This is because the datasets which are available are often purely interaction-based, meaning that they lack the necessary track-level features to implement domain-aware fairness measures. Thus, one of the preliminary steps of our work was the extensive augmentation of two publicly available datasets: LastFM \cite{LFM} and the Million Playlist Dataset \cite{MPD}. The augmentation and release of these two complementary datasets is an important contribution because it paves the way for further work in domain-aware music recommendation and alleviates the reproducibility issues often posed by the use of music datasets. Although we are limited by the number of publicly available music datasets which are compatible with our feature augmentation procedure, we believe that in selecting these two datasets, we highlight the benefits of our methodology on a broad range of settings related to music recommendation. First, these datasets encompass two important levels of recommendation: playlist (MPD) and user (LFM) based. Second, they showcase two different methods of user feedback data: implicit and explicit. MPD consists of user generated playlists meaning that its interactions consist of songs which are explicitly pronounced as relevant due to the explicit nature of a user selecting the song for their playlist. Meanwhile, LFM contains user/song interactions that are gathered by aggregating all the songs that a user clicked on (even if they did not necessarily finish or enjoy the content). Thus, these implicit interactions have no guarantee of relevance,  making the dataset more prone to noise. And, particularly in the cold start setup (see Section \ref{sec:prob_def}), this can significantly increase the difficulty of making predictions because implicit interactions are less indicative of a user's latent taste and less homogeneous in nature than that of a unified playlist. 

We augment both of our datasets to include a rich set of features scraped from Spotify API~\cite{spotipy}. To achieve this, we draw on a large body of work from the music information retrieval community (MIR) \cite{foote_content_1997, geleijsne_quest_2007}. We will publicly release the constructed datasets, the construction code, along with the code for using various feature sets, in our repository upon the publication of this paper. The details of the augmented features are as follows.
\begin{enumerate}
\vspace{-.3em}
    \item \textbf{Sonic features.} Spotify has a series of 9 proprietary features which are used to define the audio elements associated with a track. They are \textit{danceability, energy, loudness, speechiness, acousticness, instrumentalness, liveness, valence, and tempo}. Each of these features is a continuous scalar value. In order to normalize the scales, we apply $10$ leveled binning to the values.
    \item \textbf{Genre features.} We identify the primary artist associated with each collect all the genre tags associated with them. For the emebddings, we select the top $20$ and convert them to a one-hot encoding.
    \item \textbf{Track Name features.} For each song in the dataset, we extract the song title and pass it through a pre-trained language transformer model, \emph{BERT} \cite{devlin2019bert}, into an embedding of dimension $512$.
    \item \textbf{Image features.} For each song in the dataset we extract the associated album artwork. We pass this image through a pre-trained convolutional neural network, \emph{ResNet50} \cite{resnet50}, to generate an embedding of dimension $1024$.
\end{enumerate}

\subsection{Problem Setting}\label{sec:prob_settings}
The task of performing recommendation can be seen as link prediction an undirected bipartite graph. We denote such undirected bipartite graph as $G=(V, E)$. The note set $V = T \cup P$ consists of a set containing song (or track) nodes, $T$, and playlist (or user) nodes, $P$ (or $U$). The edge set $E$ are defined between a playlist $p_k$ (or user $u_k$) and a song $t_i$ if $t_i$ is contained in $p_k$ (or listened to by $u_k$). Following this setting, our goal (link prediction) is to predict whether any two song nodes $t_i, t_j \in T$ share a common parent playlist $p$.

\subsection{Reformulating Popularity Bias}\label{sec:pop_def}
\subsubsection{Defining Popularity}
\begin{figure}
  \begin{center}
   \includegraphics[width=.45\columnwidth]{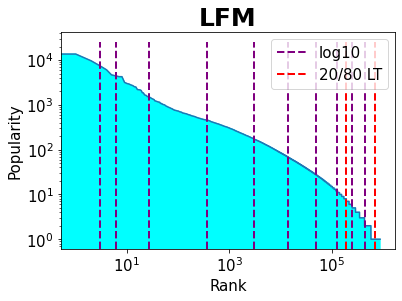}
   \includegraphics[width=.45\columnwidth]{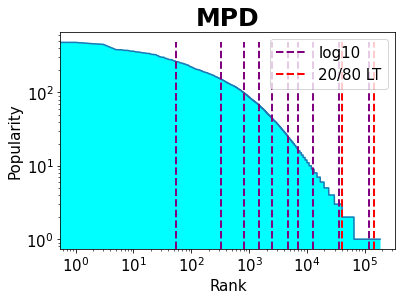}
  \end{center}
 \caption{Binning procedure for popularity definition. We show the breakdown of the bin locations for both dataset using our method as compared with the classic long tail model \cite{mansoury_graph_2021}. We can see that our  logarithmic smoothing and increased bin count allow for a more granular visualization of popularity between various item groups.}
 \label{fig:bin_breakdown}
\end{figure}
As mentioned in Section \ref{sec:pop_def_rw}, there is no true consensus within the community on how to define popularity. Here, we present a methodology which we believe allows for both the granularity and expressiveness necessary to highlight differences among various mitigation methods. Broadly, our method consists of important steps (1) logarithmic smoothing and (2) binning. In doing so, we combine the best of each methodology. Applying a logarithmic transformation to the raw values, solves the scaling issues that are caused by the extremes of the long-tail distribution. Meanwhile, binning allows us to provide aggregate statistics that highlight large scale patterns in the recommendations. And, while there are many methods which apply binning \cite{rhee_countering_2022, abdollahpouri_managing_2019, Diaz__evaluating_2020}, without the logarithmic smoothing, due to the nature of our datasets' distributions the amount of items in the bins would be unevenly distributed, leaving some bins empty. Finally, we select 10 bins based on the distribution of the datasets and the formulation of our BOOST methodology (see Section \ref{sec:r_arch}).  

We use the following steps to define our popularity setting. First, we count the number of times each song track, $t_i$ appears within the playlist (or user) training interactions such that for each $t_i$, $a_{t_i} = |\left\{p_i : t_i \in p_i \right\}|$. Then, we apply the base $10$ logarithmic smoothing to these values such that for each $t_i$, ${\rm pop}_{t_i} = \log_{10}\left(a_{t_i}\right)$. Finally, we apply binning onto these values to split them into 10 groups such that for each $t_i$,  ${\rm pop}\_{\rm bin}\left(t_i\right) \in \left\{0, \ldots, 9\right\}$ where bin $9$ has a higher popularity value than bin $0$. The visualization of this binning procedure and its comparison with the long tail method can be seen in Figure \ref{fig:bin_breakdown}. As demonstrated by our visualizations, transforming the raw values into the logarithmic space shows that the bins are filled in relatively even intervals, where, as the popularity increases, so does the number of songs included in a bin. 

We showcase the gains that our method has over the canonical long tail model in Figure \ref{fig:bin_breakdown} where we compare the positioning of our binning methodology with the classic long tail model. Furthermore, as we later show Figures \ref{fig:dset_breakdown} and \ref{fig:group_plots} our formulation of popularity is able to elucidate crucial differences among both the datasets and baseline model performances on these datasets. 
\raggedbottom
\subsubsection{Popularity Bias and Music Discovery}\label{sec:discovery_def}
In addition we formalize the inverse relationship between music discovery and popularity bias. For each song track, $t_i \in T_{OG}$, we generate a counterfactual example song, $t_i^{*} \in T_{CF}$, where everything about the features is \textit{exactly the same} and the only difference is that $t_i$ has a high degree while $t_i^{*}$ has a degree of one. We calculate the distance between an original song node, $t_i$ and its counterfactual duplicate, $t_i^{*}$. A system with high potential for musical discovery will have a low distance between the songs, showing a low popularity bias and an understanding of musical similarity. We will return to this formulation in Section \ref{sec:emb}, showing that a node's placement and degree in the graph can exacerbate the presence of popularity bias, reflecting itself in the node's learned representation. 
\subsection{Mitigating Popularity Bias Through Individual Fairness}\label{sec:r_arch}
\noindent \textbf{Ranking-based individual fairness.} 
\emph{REDRESS} is an individual fairness framework proposed by \citeauthor{dong_individual_2021} for learning fair representations in single node graphs. We extend this framework to the bipartite recommendation setting and integrate it into our popularity bias mitigation approach. In the \emph{REDRESS} setting, individual fairness requires that nodes which were similar in their initial feature space should remain similar in their learned representation embeddings \cite{dwork2011fairness}. More concretely, for each song node, $t_i$, and node pair $t_u, t_v$ in a graph $G$, similarity is defined on the basis of the cosine similarity metric, $s(\cdot, \cdot)$, as applied to either a feature $X[v] \in \Re^d$, or learned embedding set, $Z[v] \in \Re^{m}$. Applying this procedure in a pairwise fashion produces two similarity matrices. The first, or \textit{apriori similarity}, $S_G$, in which similarity is calculated on input features and the second, or \textit{learned similarity}, $S_{Z}$, in which similarity is calculated between learned embeddings generated by some GNN model, $M$. The purpose of \emph{REDRESS} is to formulate this as a ranking on the basis of similarity and differentiate on the disparity between differences in rankings of the two representational spaces. Thus, drawing on principles from learn to rank \cite{burges_lambda_2010}, each entry in these similarity matrices is re-cast as the probability that node $t_i$ is more similar to node $t_u$ than $t_v$ and transformed into an \textit{apriori probability tensor}, $P_G \in \Re^{|T| \times |T| \times |T|}$, and a \textit{learned probability tensor}, $P_Z \in \Re^{|T| \times |T| \times |T|}$. Having defined these two probability tensors, for each individual node the fairness loss, $L_{t_u,t_v}(t_i)$, can be treated the cross entropy loss applied to these probability distributions such that for an individual node, $t_i$: 
\begin{multline}
    L_{t_u,t_v}(t_i)  
    =-P_G[u,v,i] \log P_Z[u,v,i] \\ -  (1 - P_G[u,v,i]) \log (1 -  P_Z[u,v,i]) 
\end{multline}
and aggregated over all nodes $t_i \in V$ as:  
\begin{equation}\label{eq:fairness_loss}
    L_{\rm fairness} = \sum_{i}^{|T|} \sum_{u}^{|T|}\sum_{v}^{|T|} L_{t_u,t_v}(t_i) 
\end{equation}

\noindent \textbf{Individually fair music discovery.} 
The original formulation of individual fairness requires some form of domain expertise~\cite{dwork2011fairness} to determine how (dis-)similar two items are. For the music discovery domain, we use music features (see Section \ref{sec:data_aug} for exact details) as the basis for calculating cosine similarity. Thus, our \textit{apriori similarity}, $S_G$, is defined as the cosine similarity between the musical features, $X[v] \in \Re^{|T| \times 9}$, associated with song nodes. Meanwhile, our \textit{learned similarity} contains the song-level embeddings, $S_{Z} \in \Re^{|T| \times m}$, learned by PinSage. In this way, REDRESS acts as a regularizer that ensures that rank-based similarity between songs is preserved between the input and embedding space. Thus, our similarity notion is domain-aware and grounded in the essence of musical experiences: acoustics. 

To learn the embedding of songs, we follow the learning paradigm of PinSage~\cite{ying_graph_2018} and make a few deviations. Unlike \citeauthor{ying_graph_2018}, we use uniform random sampling to avoid the computational burden of calculating negative samples on our large graph and 
compensate for the potential loss of information by using focal loss~\cite{lin_2018_focal}, rather than the margin loss, to train the network. It is important to note that since the potential benefits or drawbacks of PinSage as a general recommender system are out of the scope of this paper, we do not focus on the performance gains that such a change might provide and leave the addition of various negative sampling techniques to future work.

\noindent \textbf{Bringing popularity into individual fairness.}
The \emph{REDRESS} framework does not explicitly encode any attributes of popularity in its training regimen. Thus increased visibility given of niche items comes only from innate similarities in the musical features, not explicit promotion of niche content. To extend this technique for explicitly counteracting popularity bias, we define the BOOST technique which is used to further increase the penalty on misrepresentation of items that come from diverse popularity categories. As mentioned in Section \ref{sec:pop_def}, we define 10 popularity bins by applying a logarithmic transformation and binning the degrees of a node $i$ (i.e., ${\it deg}_i$) such that  ${\rm pop}\_{\rm bin}(i) = {\rm bin} \left(\log_{10} \left({\it deg}_i\right) \right)$. This popularity bin can then be integrated with the REDRESS calculations. More formally, given the learned representation matrix, $S_Z \in \Re^{|T| \times |T|}$, we define another matrix $B$ in which 
\begin{equation}
\vspace{-.7em}
B_{ij} = |\rm pop\_bin(\textit{i}) - \rm pop\_bin(\textit{j})|
\end{equation}
Then, in the \textit{BOOST} loss formulation, in place of $S_Z$ we use $S_Z' = S_Z + B$.

\noindent \textbf{Objective function.} The representation learning objective used during training is:
\begin{equation}
\vspace{-.5em}
    L_{\rm total} = L_{\rm utility} + \gamma L_{\rm fairness}
\end{equation}
where $\gamma$ is a scalable hyperparameter which controls the focus given to fairness and can be used to select a balance between utility ($L_{\rm utility}$) and fairness ($L_{\rm fairness}$) during the second training phase. For $L_{\rm utility}$, we apply the aforementioned focal loss \cite{lin_2018_focal}. And $L_{\rm fairness}$ is Equation \ref{eq:fairness_loss} defined above.

\noindent \textbf{Generating recommendations.} 
Crucially, the PinSage architecture is only designed to learn embeddings for songs, not for playlists (or users). Thus we design our own procedure for generating playlist (or user) embeddings using the learned song embeddings. For each playlist (or user) node, $p_i$, we have a set of songs, $T(p_i) = \{t_i \in T, e_{p_i, t_i} \in E \}$, which are contained in a playlist. For a test playlist, $p_t$, we split the associated track set into two groups: 
\begin{equation*}
\vspace{-.5em}
    T(t_p) = \{t_i : t_i \in u_i\} = t_{peek} \cup t_{holdout}
\end{equation*}
such that $t_{peek}$ is a set of \textit{k} songs that are used to generate the playlist representation and $t_{holdout}$ are masked for evaluation. Thus, in order to generate a playlist (or user) embedding we define: 
\begin{equation*}
\vspace{-.5em}
    z_{p_t} = MEAN( \{ z_{t_j}: t_j \in t_{peek} \}) 
\end{equation*}
where the $z_{k} \in \Re^{1 \times d}$ are the learned representations of dimension $d$. Having learned these playlist representations, we apply cosine similarity between an individual playlist, $z_{p_t}$, and the set of songs in the database, $Z_T = \{ z_{t_j}: t_j \in T\}$, selecting the top-k items by their score. We leave further experimentation on designing user-based embeddings via the PinSage paradigm for future work.

\section{Experimental Settings}
In this section we introduce the experimental settings, defining the recommendation scenario (Section \ref{sec:prob_def}), datasets (Section \ref{sec:dataset_desc}), evaluation metrics (Section \ref{sec:metrics}), baselines (Section \ref{sec:baselines}) , and reproducibility settings (Section \ref{sec:params}) encompassed in our experimentation. 
\subsection{Recommendation Scenario} \label{sec:prob_def}
As user consumption habits have shifted away from albums and towards playlists, streaming companies have invested significant energy into the task of \textit{Automatic Playlist Continuation} and \textit{Weekly Discovery} \cite{schedl_current_2018, stanisljevic_impact_2020}. In the first task, \textit{Automatic Playlist Continuation} requires the recommender system to perform \textit{next $k$ recommendation} on a user generated playlist. Meanwhile, in \textit{Weekly Discovery}, rather than augmenting a specific playlist, an algorithm is tasked with the creation of a new  playlist based on a user's aggregated listening habits. Given our similar treatment of the playlists and users in this recommendation setting, we will use them interchangeably in our formal definition of the task. 

\begin{definition}
    Automatic Playlist Continuation/Weekly Discovery: Given a set of users $U$, or playlists, $P = P_{train} \cup P_{valid} \cup P_{test}$, and a set of songs (or tracks) $T$, our goal is to generate a set of $k$ recommendations, $R(P_{test})$. 
\end{definition}   

Following the paradigm of the cold start setting \cite{schedl_current_2018}, we extract train/validation/testing splits on the playlist level by randomly sampling without replacement such that each split trains on a distinct subset of the playlist pool. In this way, we simulate the real world situation in which new users are joining the platform and require relevant, unbiased recommendations without providing a large body of their previous interaction data. It is exactly at this junction, before a user's musical preference solidifies, that the need to mitigate popularity bias is most acute because once a majoritarian pattern has been installed in a user's embedding, it will continue to influence all further music  discovery.


    


\subsection{Datasets} \label{sec:dataset_desc}

As introduced in Section \ref{sec:data_aug}, we extensively augment two publicly available datasets, LastFM (LFM) \cite{LFM} and the Million Playlist Dataset (MPD) \cite{MPD}, with rich multi-modal track-level feature sets. Table \ref{table:dataset_stats} presents the graph statistics of both datasets. 

\begin{table}
\centering
\caption{Dataset statistics}
\footnotesize
\resizebox{\linewidth}{!}{
\begin{tabular}{|l|c|c|c|c|c|}
\hline 
Dataset &Recommendation Setting & Feedback Type & \#Users/Playlists & \#Songs & \#Artists \\
\hline 
\hline
MPD & \textit{Automatic Playlist Continuation} &Explicit& 11,100& 183,408 & 37,509\\ 
\hline 
LFM & \textit{Weekly Discovery} &Implicit & 10,267 & 890,568 & 100,638 \\ 
\hline 
\end{tabular}} 
\label{table:dataset_stats}
\end{table}
\subsection{Evaluation Metrics} \label{sec:metrics}
In addition to canonical utility metrics, we design a series of metrics to analyze the effectiveness of our debiasing methods from both a musical and fairness perspective (see Table \ref{table:metrics} for the details of the formulations). 

\subsubsection{Music Performance Metrics}
The purpose of these metrics is to broaden the scope of evaluation to include hidden positive hits. We use \emph{Artist Recall} to evaluate a system's ability to identify correct artists in a recommendation pool, an auxiliary task in music recommendation \cite{korzeniowski_artist_2021}. In addition, we design \emph{Sound Homogeneity} to capture the musical cohesiveness of the recommended songs in a playlist \cite{bontempelli_flow_2022}.  

\begin{table}
\renewcommand{\arraystretch}{2.5}
\centering

\caption{Music and fairness performance metrics. We define a ground truth set, $G$, and a recommended set, $R$, we define the set of unique artists in a playlist as $A(.)$ and the $d$-dimensional musical feature matrix associated with the tracks of a playlist as $F(.) \in \Re^{|.| \times d}$.}
\footnotesize
\resizebox{\linewidth}{!}{
\begin{tabular}{|c|c|c |}
\hline 
\textbf{Metric} & \textbf{Category} & \textbf{Formulation} \\
\hline 
\hline
\textbf{Artist Recall@100} & Music & $\frac{1}{|P_{test}|}\sum_{p \in P_{test}} \frac{1}{|A(G_p)|}{|A(G_p) \cap A(R_p)|} $ \\ 
\hline 
\textbf{Sound Homogeneity@100} &Music& $\frac{1}{|P_{test}|}\sum_{p \in P_{test}} \operatorname{cos}(F(t_i), F(t_i)) ~  \forall (t_i, t_j) \in R_p$ \\ 

\hline 
\hline 
\textbf{Artist Diversity (per playlist)} & Fairness & $\frac{1} {|P_{test}|} \sum_{p \in P_{test}} \frac{1} {|A(P)|}{|\{A(R_p)\}|} $ \\ 
\hline
\textbf{Percentage of Long Tail Items} & Fairness&$\frac{1}{|P_{test}|} \sum_{p \in P_{test}} \frac{1}{{|p|}}{|\{t_i: t_i \in R_p \cap t_i \in LT\}|}$ \\ 
\hline
\textbf{Coverage over Long Tail Items} &Fairness & $ \frac{1}{{|LT|}} {|\{t_i: t_i \in R \bigcap t_i \in |LT| \}} $\\ 
\hline
\textbf{Coverage over Artists} & Fairness& $\frac{1}{{|A|}}{|\{arid(t_i): t_i \in R|\}} $\\ 
\hline 
\end{tabular}} 
\label{table:metrics}

\end{table}
\subsubsection{Fairness Metrics}
To assess the debiasing techniques used to promote of long tail songs we draw on a series of metrics which have been previously used to evaluate the fairness performance of a model \cite{abdollahpouri_managing_2019, Patro_fairrec_2020, celma_from_2008}. Percentage metrics capture the ratio of niche to popular content that is being recommended on a playlist (or user) level. Meanwhile, \emph{coverage} looks at the aggregate sets of niche songs and artists over all recommendations. If a recommender has a high percentage but low coverage, the same niche items are being selected many times. Meanwhile, if an item has high coverage but a low percentage, the algorithm is selecting a diverse set of niche content but recommending it very rarely. The gold standard is a high value on both metrics. 

\subsection{Baselines} \label{sec:baselines} 
First, we use two naive baselines:  \textbf{(1) Features}: Instead of the learned representations, we use the raw feature vectors and \textbf{(2) MostPop}: we calculate most popular tracks in each dataset and recommend them each time. Then, we select three state of the art fairness mitigation techniques:  \textbf{(1) ZeroSum}\cite{rhee_countering_2022}: an in-processing group fairness that defines a regularization term which forces scores within negative and positive item groups to remain close. Following their original implementation, we select LGCN \cite{he_lightgcn_2020} as the backbone recommender. \textbf{ (2) MACR}\cite{wei_model_2021}: an inprocessing method which uses counterfactual estimation to denoise for the effects of popularity bias in user and item embeddings. Here too, Following their original implementation, we select LGCN \cite{he_lightgcn_2020} as the backbone recommender, and \textbf{(3) Smooth xQuAD}\cite{abdollahpouri_managing_2019}: a post-processing method that reranks recommendations to improved diversity. 

\subsection{Parameter Settings \& Reproducibility} \label{sec:params}
Each of the baseline methods was tested with learning rates  $\sim (0.01, 0.0001)$, embedding sizes of $[10, 24, 64, 128]$ and batch sizes of $[256, 512, 1024]$. For the values in the tables below, each stochastic method was run $5$ times and averaged. All details and further hyperparameter settings can be found on our GitHub repository \footnote{preliminary version:~ \url{https://anonymous.4open.science/r/RecSys23-9B7F/README.md}}.  

\section{Results} \label{sec:results}
In this section we present the results of our experimentation. First, we show the connections between individual fairness, popularity bias, and music discovery in the graph domain. Then, we evaluate our method, comparing with a series of the debiasing benchmarks. 

\raggedbottom
\subsection{RQ1: How does incorporating individual fairness improve the mitigation of popularity bias and facilitate music discovery?}\label{sec:emb}
To showcase the performance of our algorithm in the discovery setting and motivate the need for individual fairness in the mitigation of popularity bias, we draw on the definition of music discovery presented in Section \ref{sec:discovery_def} by 
evaluating the effects of popularity bias on learned representations of popular and unpopular songs. 

\begin{figure}
\centering
\includegraphics[width=.49\columnwidth]{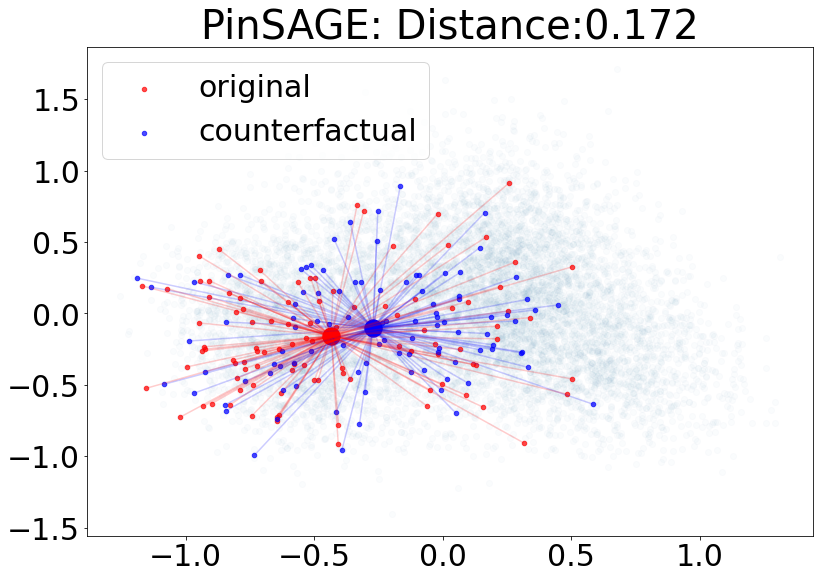}
\includegraphics[width=.49\columnwidth]{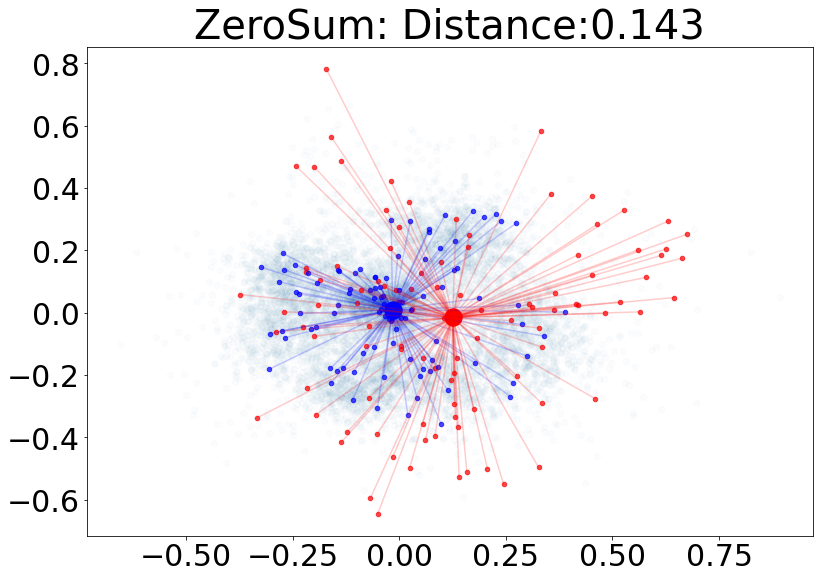}
\includegraphics[width=.49\columnwidth]{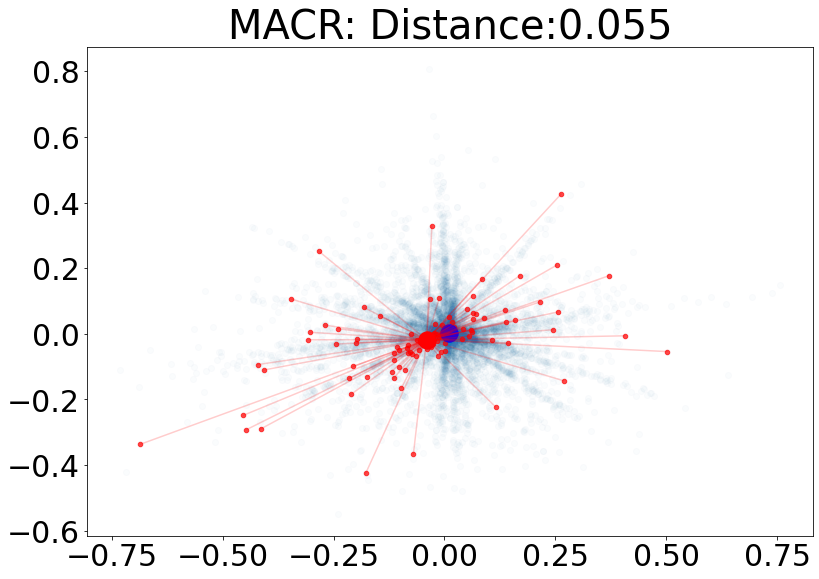}
\includegraphics[width=.49\columnwidth]{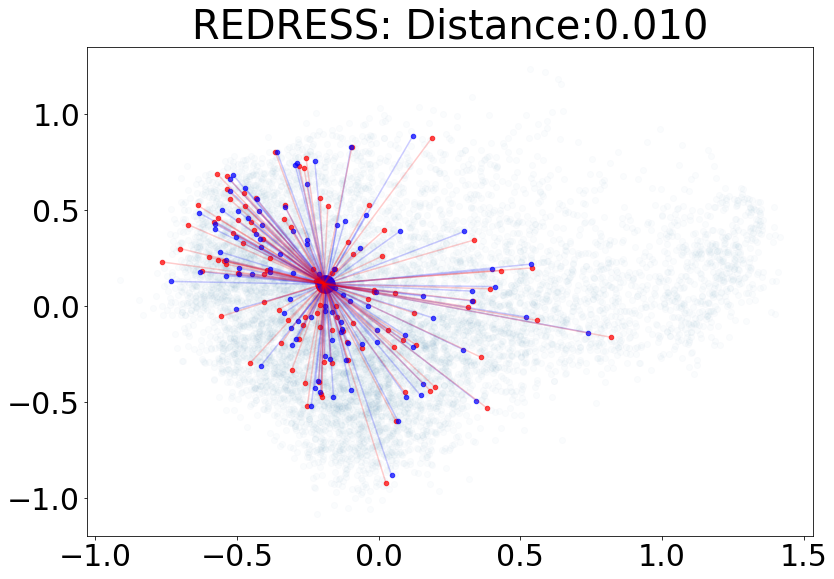}
\includegraphics[width=.49\columnwidth]{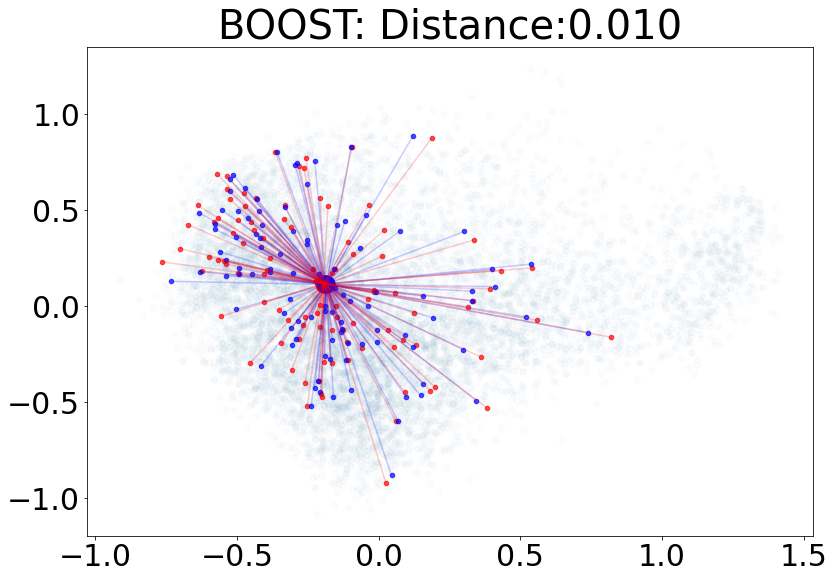}
\caption{Simulating Popularity Bias:  We select $100$ of the most popular songs in MPD \cite{schedl_current_2018}, duplicate their node level features, add them with new song track ids, and connecting them to one randomly selected playlist. Then, we analyze the distances between the embedding group centroids. We find that REDRESS and BOOST have the lowest distance between the original, popular and duplicated, unpopular track groups, showing the least amount of popularity bias.}
\label{fig:cf_emb}
\end{figure}

To simulate a situation of maximal popularity bias, we consider the hypothetical example in which extremely popular songs are reversed to become unpopular and measure the effects of degree on their learned representations. From a discovery perspective, the purpose of this simulation is to imagine the most popular song by a listener's favorite artist before it became popular. Our simulation aims to approximate how likely it is that they have discovered the song in relation to its musical attributes, with and without debiasing for the effects of popularity. More formally, for each song track, $t_i \in T_{OG}$, we generate a counterfactual example song, $t_i^{*} \in T_{CF}$, where everything about the features is \textit{exactly the same} and the only difference is that $t_i$ appears in many playlists while $t_i^{*}$ appears only once. We augment the original dataset to include these counterfactual songs, $T = T_{OG} \bigcup T_{CF}$. Then, we use five methods to learn the item level representations: one baseline recommender, PinSage, and four bias mitigation methods, ZeroSum \cite{rhee_countering_2022}, MACR \cite{wei_model_2021}, REDRESS, and BOOST. We apply 2-dimensional PCA to each embedding set and analyze the Euclidean distance between the centroids of original track embeddings, $\bar{T}_{OG}$, and counterfactual track embeddings, $\bar{T}_{CF}$. Due to the size of our dataset, we run this metric using the 100 most popular tracks in the MPD dataset and leave further exploration of this phenomenon for future work.  

As shown in Figure \ref{fig:cf_emb}, we find that all fairness interventions decrease the distance between the two centroids. Furthermore, as the granularity of fairness increases, the distance between the centroids of learned representations decreases. For example, PinSage, which has no mitigation of popularity bias, has the largest distance of $0.172$. ZeroSum \cite{rhee_countering_2022}, which considers group fairness, decreases the distance to $0.143$, MACR \cite{wei_model_2021}, which uses counterfactual estimation, shrinks to $0.055$. Finally, our methods, REDRESS and BOOST are able to achieve both the lowest distance and the correct orientation between the two embedding spaces. 

In these results, we see that the domain-awareness of our methodology, which enables it to understand musical similarity between items, allows it to be robust to the effects of popularity bias on a learned song embedding. Thus, in the setting of musical discovery, it is able to uncover proximity between items which are musically coherent even if they are not necessarily of similar popularity status. And, in doing so, we build representations that are complex, expressive, and effective for music recommendation.  
\subsection{RQ2: How does our individual fairness approach compare to existing methods for mitigating popularity bias?}
\begin{table*} 
\centering
\caption{Comparison between all methods.Note: We use bold highlights to represent the best performance within a column for each of the datasets. The $p$ values of this table are calculated by applying the Wilcoxon signed-rank test \cite{rey_wilcoxon-signed-rank_2011} to results between PinSage and BOOST. As you can see, the \emph{BOOST} method achieves the best performance along all Fairness metrics when compared with the other debiasing benchmarks.}
\label{table:baseline_comp}
\footnotesize
\resizebox{\linewidth}{!}{
\begin{tabular}{|ll|cc|cc|cccc|}
\hline && \multicolumn{2}{|c|}{\textbf{Classic}}  
& \multicolumn{2}{|c|}{\textbf{Music}} 
&  \multicolumn{4}{|c|}{\textbf{Fairness}} \\ 
\hline 
\textbf{Data} & \textbf{Model} & 
      Recall@100 & 
      NDCG@100 & 
       Artist Recall@100 & 
      Flow &
      Diversity &
      \%LT &
      LT Cvg &
      Artist Cvg \\
\hline 
\hline 
\multirow{9}{4em}{\textbf{MPD}}&\textbf{Features}  & 
    0.041 & 
    0.073 & 
    0.073 &
    0.900 & 
    0.841 & 
    \textbf{0.588} & 
    0.022 &  
    0.073 \\ 
&\textbf{MostPop}  & 
    0.044 & 
    0.048 & 
    0.141 &
    0.908 & 
    0.680 & 
    0.0 & 
    0.0 &  
    0.001 \\ \cline{2-10}

&\textbf{LightGCN}&
      \textbf{0.106 $\pm$ 0.004} & 
      0.119 $\pm$ 0.004&
       \textbf{0.272 $\pm$ 0.011} & 
      0.905 $\pm$ 0.000& 
      0.672 $\pm$ 0.025&
      0.002 $\pm$ 0.000& 
      0.000 $\pm$ 0.000& 
      0.025 $\pm$ 0.001\\ 
& \textbf{PinSage} & 
    0.068 $\pm$ 0.002 & 
     \textbf{0.144 $\pm$ 0.003} & 
    0.139 $\pm$ 0.003 &
    0.931 $\pm$ 0.001 & 
    0.707 $\pm$ 0.003 & 
    0.476 $\pm$ 0.002 &  
    0.032 $\pm$ 0.000 &  
    0.105 $\pm$ 0.000 \\ \cline{2-10}
    
&\textbf{ZeroSum}   &  
    0.044 $\pm$ 0.002& 
    0.043 $\pm$ 0.002& 
    0.220 $\pm$ 0.008 &
    0.904 $\pm$ 0.001& 
    0.765 $\pm$ 0.013& 
    0.000 $\pm$ 0.003& 
    0.000 $\pm$ 0.000&  
    0.048 $\pm$ 0.002 \\

&\textbf{xQuAD}  & 
    0.064 $\pm$ 0.005 & 
    0.104 $\pm$ 0.006 & 
    0.135 $\pm$ 0.013&
    0.927 $\pm$ 0.004& 
    0.703 $\pm$ 0.059& 
    0.226 $\pm$ 0.001& 
    0.017 $\pm$ 0.000&  
    0.098 $\pm$ 0.004 \\ 

&\textbf{MACR}  & 
    0.028 $\pm$ 0.014& 
    0.030 $\pm$ 0.015& 
    0.149 $\pm$ 0.022&
    0.902 $\pm$ 0.002& 
    0.831 $\pm$ 0.034& 
    0.019 $\pm$ 0.006& 
    0.000 $\pm$ 0.001&  
    0.011 $\pm$ 0.003 \\ \cline{2-10}
& \textbf{REDRESS}  & 
    0.045 $\pm$ 0.002 & 
    0.100 $\pm$ 0.003 & 
    0.162 $\pm$ 0.004 &
    0.969 $\pm$ 0.032& 
    0.829 $\pm$ 0.001& 
    0.504 $\pm$ 0.003& 
    0.036 $\pm$ 0.004 &  
    0.117 $\pm$ 0.000 \\ 
&  \textbf{BOOST} & 
    0.020 $\pm$ 0.004 & 
    0.047 $\pm$ 0.003& 
    0.137 $\pm$ 0.002 &
    \textbf{0.979 $\pm$ 0.000} & 
    \textbf{0.899 $\pm$ 0.002}& 
    \textbf{0.522 $\pm$ 0.001}& 
    \textbf{0.037 $\pm$0.003}&  
    \textbf{0.125 $\pm$0.000} \\
\hline 
&  \textit{p values} & 
    4.408083e-16	 & 
    1.768725e-19& 
    0.727897	 &
    3.751961e-61 & 
    1.168816e-29 & 
    0.000596& 
    -&  
    -  \\
\hline
\hline 

\multirow{9}{4em}{\textbf{LFM}}&\textbf{Features}  & 
    0.033 & 
    0.037 & 
    0.041 &
    0.996 & 
    0.919 & 
    0.486 & 
    0.005 &  
    0.034 \\
 &\textbf{MostPop}  & 
    0.015 & 
    0.011 & 
    0.046 &
    0.926 & 
    0.600 & 
    0.000 & 
    0.000 &  
    0.001 \\ \cline{2-10}
&\textbf{LightGCN} &
      0.026 $\pm$ 0.001	 & 
      0.023	 $\pm$ 0.001	&
      0.068 $\pm$ 0.001	& 
      0.998$\pm$ 0.000 & 
      0.505 $\pm$ 0.012&
      0.000 $\pm$ 0.000& 
      0.000 $\pm$ 0.000& 
      0.003 $\pm$ 0.001	\\ 
& \textbf{PinSage} & 
    \textbf{0.064 $\pm$ 0.001} & 
    \textbf{0.095 $\pm$ 0.002} & 
    \textbf{0.077  $\pm$ 0.002} &
    0.969 $\pm$ 0.000 & 
    0.775 $\pm$ 0.003 & 
    0.437 $\pm$ 0.001 &  
    0.008  $\pm$ 0.000 &  
    0.053  $\pm$ 0.001 \\ \cline{2-10}
&\textbf{ZeroSum}  & 
    0.001 $\pm$ 0.003& 
    0.001 $\pm$ 0.001& 
    0.045 $\pm$ 0.004&
    0.996 $\pm$ 0.008& 
    0.866 $\pm$ 0.000& 
    0.007 $\pm$ 0.000& 
    0.000 $\pm$ 0.000&  
    0.032 $\pm$ 0.001 \\
    
& \textbf{xQuAD} & 
    0.055 $\pm$ 0.001 & 
    0.064 $\pm$ 0.001 & 
    0.068 $\pm$ 0.002 &
    0.998 $\pm$ 0.000 & 
    0.801 $\pm$ 0.008 & 
    0.212 $\pm$ 0.000 & 
    0.004 $\pm$ 0.000 &  
    0.053 $\pm$ 0.001 \\

&\textbf{MACR}  & 
    0.014 $\pm$ 0.001& 
    0.014 $\pm$ 0.001& 
    0.049 $\pm$ 0.007&
    0.996 $\pm$ 0.003& 
    0.777 $\pm$ 0.050& 
    0.002 $\pm$ 0.004& 
    0.000 $\pm$ 0.000&  
    0.001 $\pm$ 0.000 \\ \cline{2-10}
    
&\textbf{REDRESS} & 
    0.038 $\pm$ 0.002 & 
    0.053 $\pm$ 0.004 & 
    0.057 $\pm$ 0.001 &
    0.998 $\pm$ 0.002& 
    0.862 $\pm$ 0.004& 
    0.451 $\pm$ 0.000& 
    0.008 $\pm$ 0.002&  
    0.056 $\pm$ 0.000 \\


 &\textbf{BOOST}  & 
    0.005 $\pm$ 0.001&  
    0.007 $\pm$ 0.001& 
    0.029 $\pm$ 0.002&
    \textbf{0.999 $\pm$ 0.000} & 
    \textbf{0.941 $\pm$ 0.003} & 
    \textbf{0.498 $\pm$ 0.006} & 
    \textbf{0.010 $\pm$ 0.000} &  
    \textbf{0.068 $\pm$ 0.001} \\ 
\hline 
&  \textit{p values} & 
    5.696989e-08	 & 
    1.179627e-15& 
    1.914129e-07&
    0.001408 & 
    1.112495e-34&  
    2.477700e-11& 
    -&  
    - \\
\hline 
\end{tabular}}
\end{table*}

\begin{figure} 
\centering
\includegraphics[width=\columnwidth]{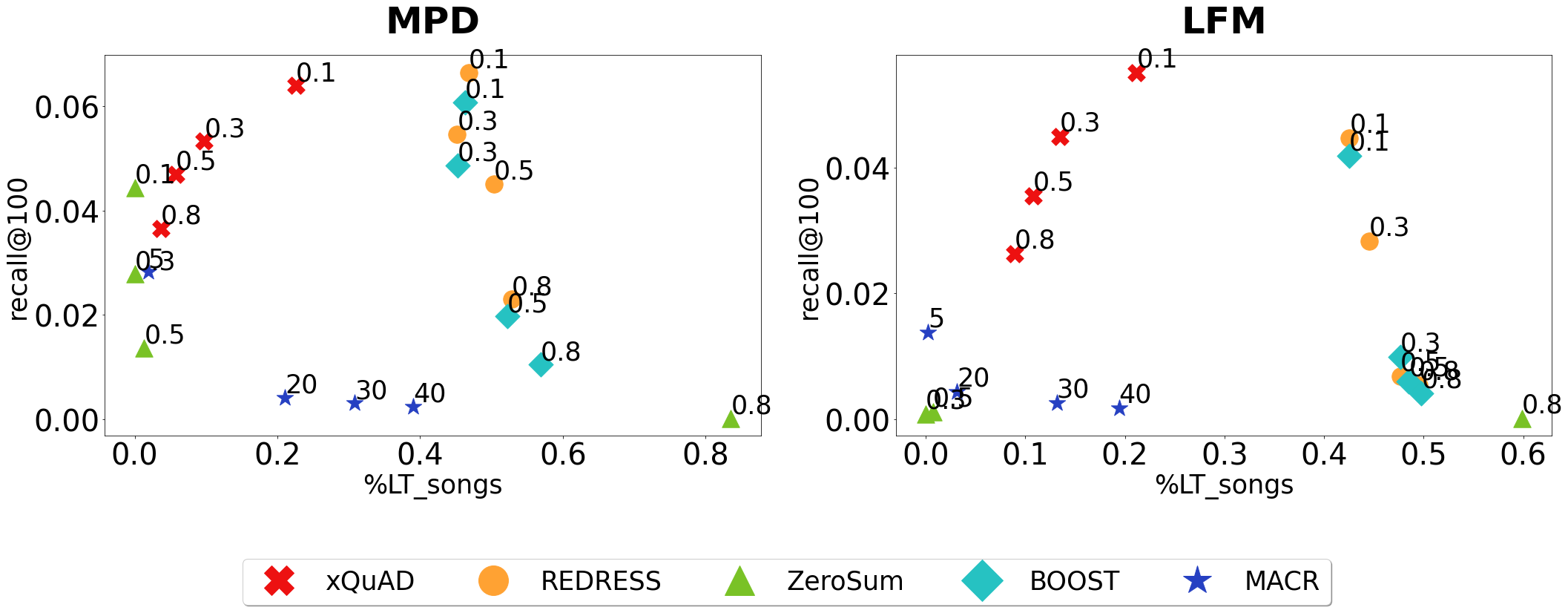}
\caption{\textbf{Plots showing the effects of fairness intervention method on performance/fairness metrics}: we show the effects of tuning a hyperparameter to balance fairness and performance in each of our fairness methods, we explore the entire range of the values and report the trade-off that increasing this fairness intervention can have. Note, that REDRESS has the best robustness with respect to balancing recall and \%LT. }
\label{fig:pareto_plots}
\end{figure}

In Table \ref{table:baseline_comp} we show a side by side comparison of the various recommendation and debiasing methods. We apply the Wilcoxon signed-rank test \cite{rey_wilcoxon-signed-rank_2011} to the BOOST, PinSage results to assess the statistical significance of our method's performance. In Table \ref{table:baseline_comp}, we select only the best hyperparameter results for each method. However, for each fairness baseline, there is a hyperparameter which tunes the balance between fairness intervention and performance, $\gamma$. Thus, we present Figure \ref{fig:pareto_plots} to show the ranges of these hyperparameter values. 

\begin{figure}
\centering
\includegraphics[width=.45\columnwidth]{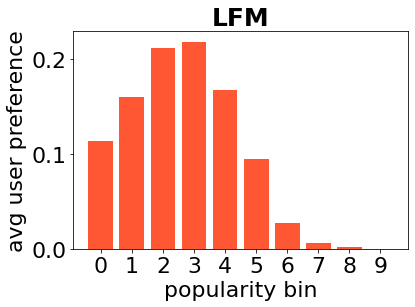}
\includegraphics[width=.45\columnwidth]{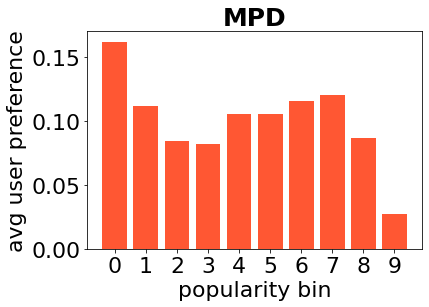}
\caption{\textbf{Dataset Breakdown by Long Tail Definition}: We show visualizations of the user preferences indicated in the training set for each of the datasets used in evaluation. Using our formulation of popularity we can see that the two datasets have different distributions of popularity in their training data which, in turn, helps explain fairness/performance tradeoffs.} 
\label{fig:dset_breakdown}
\end{figure}

\noindent \textbf{Analyzing Debiasing Performance}: 
First, we look at the comparison between the backbone recommender systems and their debiasing counterparts. Within the greater fairness community it is typical to see a trade-off between recommendation utility and the effectiveness of a debiasing technique \cite{herlocker_evaluating_2004}. And, indeed, in our experiments we witness such a trade-off. For example, evaluating the columns of \textit{Recall} and \textit{NDCG} on Table \ref{table:baseline_comp} we can see that both recommender systems outperform their debiasing counterparts. This can be largely attributed to the formulations of the utility metrics, the interaction with the debiasing objectives, and the underlying distribution of the datasets. Since the premise of the canonical recommendation utility metrics is to reward a system that can accurately recover the exact tracks a user liked, any attempts to promote long tail content that wasn't originally listened to is penalized, even if it isn't truly indicative of a user's underlying taste. We note that the trade-off is significantly more severe in the LFM dataset as opposed to the MPD dataset. As shown in Figure \ref{fig:dset_breakdown}, this can be attributed to the underlying distribution of a datasets. Where MPD has a training set which contains a significant portion of interactions on the lower end of the popularity spectrum, LFM skews towards higher popularity. Thus, if evaluating using utility metrics that penalize mistakes on the track-level prediction, even if a system is selecting musically coherent and relevant content, this trade-off becomes inevitable. Indeed, within the music recommendation community, there have been several works suggesting that this trade-off, though present in offline testing, doesn't necessarily carry over into online testing \cite{castells_offline_2022, herlocker_evaluating_2004}. Thus, to provide a deeper analysis of our debiasing methodology, we present two musically relevant metrics, \textit{Artist Recall} and \textit{Flow}. As explained in Section \ref{sec:metrics} and detailed in Table \ref{table:metrics}, the \textit{Artist Recall} metric evaluates the recommender systems ability to identify correct artist-level recommendations and the \textit{Flow} evaluates the overall homogeneity of the selected music. In particular, \textit{Flow} plays an important role in the music discovery task because studies have indicated that users are drawn to homogeneous listening suggestions when engaging with algorithmic curation \cite{bontempelli_flow_2022, herlocker_evaluating_2004}. As we can see in both datasets, REDRESS and BOOST consistently achieve the highest \textit{Flow}. This is because, by harnessing musical features and in our debiasing technique, our method generates representations that are indicative of musical similarity, which affects the downstream musical similarity of the recommendations it generates. Meanwhile, looking at the \textit{Artist Recall} columns, we can see a much less significant drop (or, in the case of MPD an increase) in the performance between backbone recommender models and their debiasing counterparts. Crucially, if we consider the implications of such a debiasing technique on a user who's taste skews towards popular music, high performance on these metrics means that our debiasing methods' awareness of musical similarity will enable it to maintain the stylistic elements that a user is drawn to while simultaneously promoting niche content.  


Next, we compare the performance among the various fairness promotion methods. Looking at the columns of \textit{recall} and \textit{ndcg} on Table \ref{table:baseline_comp}, we can see that, as expected, xQuAD \cite{abdollahpouri_managing_2019} which is a re-ranking method is able to preserve the highest utility. However, we note that among the remaining methods, REDRESS is able to achieve the second highest utility. Meanwhile, if we look at the fairness metrics, we can see that REDRESS and BOOST are the highest performing methods. In particular, looking at the columns for \textit{\%LT} and \textit{LT Cvg}, we can see that REDRESS is noticeably better than the other methods and BOOST is able to improve on its performance. Crucially, our method is able to have high values in both coverage and percentage of long tail items meaning that REDRESS/BOOST is not just prioritizing niche items but also choosing a diverse selection from this category. We attribute the relative gains of REDRESS and BOOST to their ability to integrate musical features into their fairness mitigation because they are able to select not just niche items, but also musically relevant ones for recommendation. Finally, our method's ability to interpolate between these two perspectives of content and consumption patterns, shows that REDRESS/BOOST is able to recommend similar ratios of niche items compared to the bare features while having significantly better performance.

\noindent \textbf{Hyperparameter Sensitivity}:
While the results in Table \ref{table:baseline_comp} are compared among the best hyperparameter tuning that balances between utility and fairness, we also present Figure \ref{fig:pareto_plots} where we show the balance between \textit{\%LT} and \textit{recall} along the range of each method's hyperparameter value. For example, xQuAD, ZeroSum, REDRESS and BOOST all have ranges that scale between $(0.1, 0.9)$ and MACR has a value somewhere along $(0, 45)$. As we can see, for any value of hyperparameter along the various methods, REDRESS and BOOST are able to outperform the collection of benchmarks. Given these results, we conclude that our BOOST approach is able to achieve the most effective debiasing performance while REDRESS is able to achieve the most balanced performance.

\begin{figure} 
\centering
\includegraphics[width=\columnwidth]{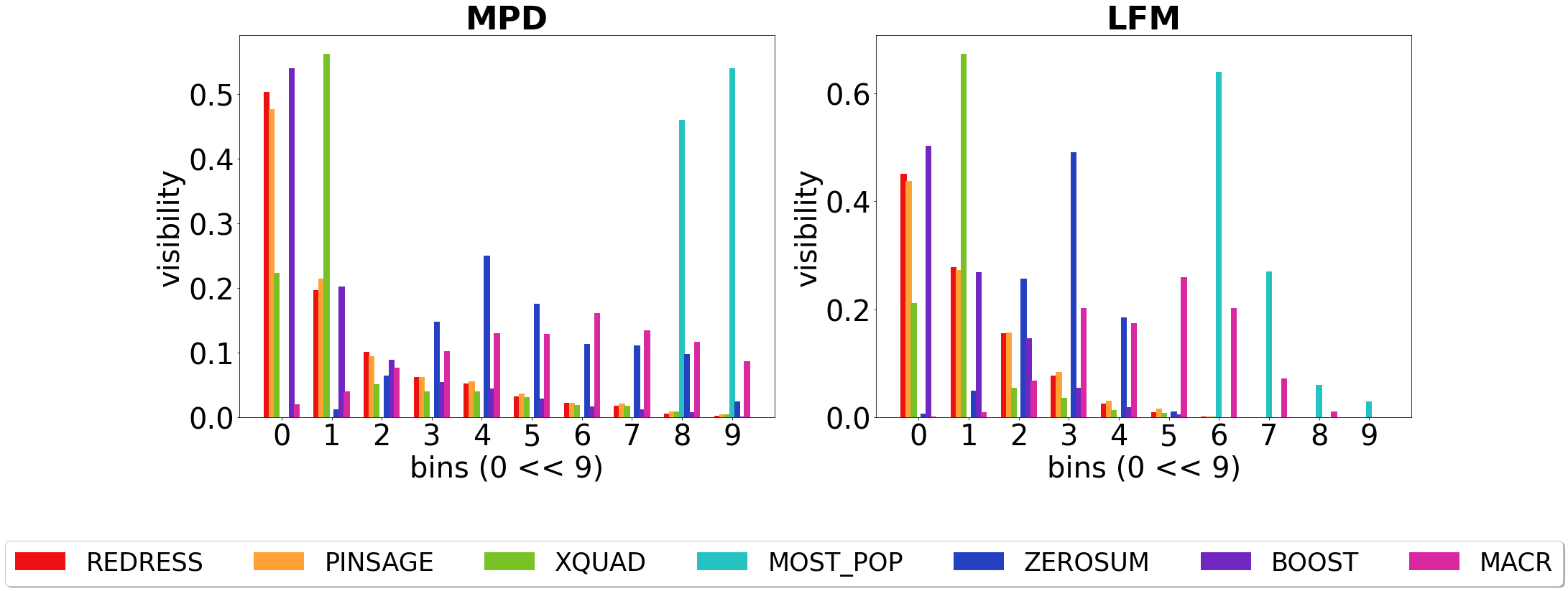}
\caption{\textbf{Group By Group Analysis of Recommendations}: we look at a breakdown of the recommendations for each dataset. We define visibility as the number of times an item from this group appears in the recommendations normalized by the total number of items in the recommendation lists. Bins are defined using the methodology of Section \ref{sec:pop_def} where bin $0$ has the lowest popularity.}
\label{fig:group_plots}
\end{figure}

\noindent\textbf{Popularity Definition}:  As we can see in Figure \ref{fig:group_plots} the definition of popularity plays a significant role in the model selection method \textit{especially} in the case where user preferences encoded in the training data skew towards popular items. In particular, using a less granular definition for popularity bins can synthetically inflate the performance of $\%LT$ and $LT Cvg$. For example, we can see that methods like xQuAD and ZeroSum are selecting a majority of their items from bins 1,2 or 3. Using a classical long tail methodology, these differences would not be as visible, masking distinctions among the baselines' fairness.

\section{Conclusion}
In this work, we address the problem of mitigating popularity bias in music recommendation. Starting from the perspective of discovery and how it relates to algorithmic curation, we consider the effects of popularity bias on users' ability to discover novel and relevant music. On the basis of this motivation, we highlight the intrinsic ties between popularity bias and individual fairness on both song and artist levels. We ground our individual fairness notion in the music domain, presenting a method to mitigate popularity bias through fine tuning of representation learning via musical similarities. We perform extensive evaluation on two music datasets showing the improvements of our domain aware method in comparison with three state of the art popularity bias mitigation techniques. We hope that these promising findings showcase the importance of developing domain aware methods of mitigating popularity bias in addition to domain agnostic options. 
\newpage
\bibliographystyle{ACM-Reference-Format}

\bibliography{ref}

\section{Ethical Considerations}
In this work we attempt to promote ethical recommendation by addressing the problem of popularity bias in music recommendation. However, there are a few important caveats to the work presented above. Crucially, our work is influenced by larger movements within the music streaming and information retrieval communities that come with important ethical considerations. First, due to our reliance on publicly available data, we are limited in our exploration of multi-cultural, and specifically non-Western musical content. The under representation of such content has long been an issue within the music communities and, concretely in our setting, it manifests itself in cutting off the flow of financial capital towards under-represented artists. Second, due to the lack of transparency regarding the calculation of Spotify's metadata, we are unable to assess whether the musical features distributed by the platform (and used in our work) are disenfranchising content from non-Western musical traditions. Finally, the construction of user bases can also have very significant effects on how connections between artists are formed. Thus, it is equally important to note that the ground truth data that we are using to perform recommendation is most likely very dominated by North American and European listeners. We hope that in the future, greater collaboration between industry and academia can foster the necessary transparency to address these issues more concretely. 


\end{document}